\documentclass[twoside,twocolumn,9pt]{article}
\usepackage{extsizes}
\usepackage[super,sort&compress,comma]{natbib}
\usepackage[version=3]{mhchem}
\usepackage[left=1.5cm, right=1.5cm, top=1.785cm, bottom=2.0cm]{geometry}
\usepackage{balance}
\usepackage{times,mathptmx}
\usepackage{sectsty}
\usepackage{graphicx}
\usepackage{lastpage}
\usepackage[format=plain,justification=justified,singlelinecheck=false,font={stretch=1.125,small,sf},labelfont=bf,labelsep=space]{caption}
\usepackage{float}
\usepackage{fancyhdr}
\usepackage{fnpos}
\usepackage[english]{babel}
\usepackage{array}
\usepackage{droidsans}
\usepackage{charter}
\usepackage[T1]{fontenc}
\usepackage[usenames,dvipsnames]{xcolor}
\usepackage{setspace}
\usepackage[compact]{titlesec}
\usepackage{epstopdf}%This line makes .eps figures into .pdf - please comment out if not required.

\definecolor{cream}{RGB}{222,217,201}

\begin{document}

\pagestyle{fancy}
\thispagestyle{plain}
\fancypagestyle{plain}{

%%%HEADER%%%
\fancyhead[C]{}
\fancyhead[L]{\hspace{0cm}\vspace{1.5cm}}
\fancyhead[R]{\hspace{0cm}\vspace{1.7cm}}
\renewcommand{\headrulewidth}{0pt}
}
%%%END OF HEADER%%%

%%%PAGE SETUP - Please do not change any commands within this section%%%
\makeFNbottom
\makeatletter
\renewcommand\LARGE{\@setfontsize\LARGE{15pt}{17}}
\renewcommand\Large{\@setfontsize\Large{12pt}{14}}
\renewcommand\large{\@setfontsize\large{10pt}{12}}
\renewcommand\footnotesize{\@setfontsize\footnotesize{7pt}{10}}
\makeatother

\renewcommand{\thefootnote}{\fnsymbol{footnote}}

\makeatletter
\renewcommand\@biblabel[1]{#1}
\renewcommand\@makefntext[1]%
{\noindent\makebox[0pt][r]{\@thefnmark\,}#1}
\makeatother
\renewcommand{\figurename}{\small{Fig.}~}
\sectionfont{\sffamily\Large}
\subsectionfont{\normalsize}
\subsubsectionfont{\bf}
\setstretch{1.125} %In particular, please do not alter this line.
\setlength{\skip\footins}{0.8cm}
\setlength{\footnotesep}{0.25cm}
\setlength{\jot}{10pt}
\titlespacing*{\section}{0pt}{4pt}{4pt}
\titlespacing*{\subsection}{0pt}{15pt}{1pt}
%%%END OF PAGE SETUP%%%

%%%FOOTER%%%
\fancyfoot{}
\fancyfoot[LO,RE]{\vspace{-7.1pt}}
\fancyfoot[CO]{\vspace{-7.1pt}\hspace{13.2cm}}
\fancyfoot[CE]{\vspace{-7.2pt}\hspace{-14.2cm}}
\fancyfoot[RO]{\footnotesize{\sffamily{1--\pageref{LastPage} ~\textbar  \hspace{2pt}\thepage}}}
\fancyfoot[LE]{\footnotesize{\sffamily{\thepage~\textbar\hspace{3.45cm} 1--\pageref{LastPage}}}}
\fancyhead{}
\renewcommand{\headrulewidth}{0pt}
\renewcommand{\footrulewidth}{0pt}
\setlength{\arrayrulewidth}{1pt}
\setlength{\columnsep}{6.5mm}
\setlength\bibsep{1pt}
%%%END OF FOOTER%%%

%%%FIGURE SETUP - please do not change any commands within this section%%%
\makeatletter
\newlength{\figrulesep}
\setlength{\figrulesep}{0.5\textfloatsep}

\newcommand{\topfigrule}{\vspace*{-1pt}%
\noindent{\color{cream}\rule[-\figrulesep]{\columnwidth}{1.5pt}} }

\newcommand{\botfigrule}{\vspace*{-2pt}%
\noindent{\color{cream}\rule[\figrulesep]{\columnwidth}{1.5pt}} }

\newcommand{\dblfigrule}{\vspace*{-1pt}%
\noindent{\color{cream}\rule[-\figrulesep]{\textwidth}{1.5pt}} }

\makeatother
%%%END OF FIGURE SETUP%%%

%%%TITLE, AUTHORS AND ABSTRACT%%%
\twocolumn[
  \begin{@twocolumnfalse}
\vspace{3cm}
\sffamily
\begin{tabular}{m{4.5cm} p{13.5cm} }

 & \noindent\LARGE{\textbf{Doping anatase TiO$_2$ with group V-b and VI-b transition metal atoms : a hybrid functional first-principles study}}
\\%Article title goes here instead of the text "This is the title"
\vspace{0.3cm} & \vspace{0.3cm} \\

 & \noindent\large{Masahiko Matsubara,\textit{$^{a,c}$} Rolando Saniz,\textit{$^{b}$} Bart Partoens,\textit{$^{b}$} and Dirk Lamoen$^{\ast}$\textit{$^{a}$}}
 \\%Author names go here instead of "Full name", etc.

 & \noindent\normalsize{We investigate the role of transition metal atoms of
 group V-b (V, Nb, Ta) and VI-b (Cr, Mo,
W) as n- or p-type dopants in anatase TiO$_2$ using thermodynamic
principles and density functional theory with the
Heyd-Scuseria-Ernzerhof HSE06 hybrid functional. The HSE06
functional provides a realistic value for the band gap, which
ensures a correct classification of dopants as shallow or deep
donors or acceptors. Defect formation energies and thermodynamic
transition levels are calculated taking into account the
constraints imposed by the stability of TiO$_2$ and the solubility
limit of the impurities.
%and the thermodynamic stability of the occurring
%Ti and the transition metal oxides.
Nb, Ta, W and Mo are identified as shallow donors. Although W
provides two electrons, Nb and Ta show a considerable lower
formation energy, in particular under O-poor conditions. Mo
donates in principle one electron, but under specific conditions
can turn into a double donor. V impurities are deep donors and Cr
shows up as an amphoteric defect, thereby acting as an electron
trapping center in n-type TiO$_2$ especially under O-rich
conditions. A comparison with the available experimental data
yields excellent agreement.} \\%The abstrast goes here instead of the text "The abstract should be..."

\end{tabular}

 \end{@twocolumnfalse} \vspace{0.6cm}

  ]
%%%END OF TITLE, AUTHORS AND ABSTRACT%%%

%%%FONT SETUP - please do not change any commands within this section
\renewcommand*\rmdefault{bch}\normalfont\upshape
\rmfamily
\section*{}
\vspace{-1cm}

%%%FOOTNOTES%%%

\footnotetext{\textit{$^{a}$~EMAT , Departement Fysica, Universiteit Antwerpen, Groenenborgerlaan 171, B-2020 Antwerpen, Belgium}}
\footnotetext{\textit{$^{b}$~CMT, Departement Fysica, Universiteit Antwerpen, Groenenborgerlaan 171, B-2020 Antwerpen, Belgium }}
\footnotetext{\textit{$^{b}$~ECE, Boston University, 8 Saint Mary's Street, Boston, MA 02215, USA}}

%Please use \dag to cite the ESI in the main text of the article.
%If you article does not have ESI please remove the the \dag symbol from the title and the footnotetext below.
%\footnotetext{\dag~Electronic Supplementary Information (ESI) available: [details of any supplementary information available should be included here]. See DOI: 10.1039/b000000x/}
%additional addresses can be cited as above using the lower-case letters, c, d, e... If all authors are from the same address, no letter is required

\footnotetext{$\ast$~Corresponding author:Dirk.Lamoen@uantwerpen.be}

%%%END OF FOOTNOTES%%%

%%%MAIN TEXT%%%%
\section{Introduction\label{sec:intro}}
TiO$_{2}$ is an important material in the field of renewable
energy applications, and well-known as an efficient photocatalyst
used e.g. for the hydrogen production by water splitting.\cite{hashimoto05} In addition, TiO$_{2}$, more specifically the
anatase phase, is suitable for producing novel materials by doping
with transition metal (TM) atoms, because of their high solubility
in this phase.\cite{gracia04} Anatase itself is a wide band gap
($\sim$3.4 eV\cite{tangh95}) semiconductor, which is low cost,
nontoxic and chemically stable. On doping with group V-b elements
Nb\cite{furubayashi05,furubayashi06}, Ta\cite{hitosugi05} or
with W (group VI-b)\cite{sathasivam15} it becomes a good
transparent conducting oxide (TCO).

Thus, TM-doped anatase is considered to be one of the promising
candidates as an effective TCO to replace the currently most used
but expensive indium-tin-oxide in the near future.
The novelty of doped anatase is not limited to TCOs. When doped
with other TM atoms, it exhibits different properties e.g.
% Meanwhile,
when anatase is doped with Co, it becomes a dilute magnetic
semiconductor (DMS) showing ferromagnetism at room
temperature.\cite{matsumoto01,janisch06}
Furthermore, V-doped anatase is a potential DMS with a giant
magnetic moment\cite{hongnh04} and Cr-doped anatase is also
shown to become a DMS with a magnetization of $\sim$ 0.6
$\mu_B$/Cr atom.\cite{droubay05} Doping with Mo improves the
photocatalytic activity of TiO$_2$\cite{stengl, khan,wang} and
increases the conductivity.\cite{boen}
%

%Osorio-Guillen \emph{et al.} studied theoretically the electronic
%behavior of group V-b TM (V, Nb and Ta) doped anatase using
%density functional theory (DFT) within the generalized gradient
%approximation (GGA) and showed that the atomic $d$ orbital energy
%of the doped TM atom is the deciding factor for these materials to
%become a TCO or DMS.~\cite{osorio-guillen08}

In this paper we perform a systematic analysis of the dopant
characteristics of elements from group V-b (V, Nb, Ta) and group
VI-b (Cr, Mo, W) in anatase TiO$_2$ based on thermodynamic
principles and density functional theory (DFT).
%with the
%Heyd-Scuseria-Ernzerhof (HSE) hybrid functional~\cite{heyd03,
%heyd06}.
% to reproduce the correct band gap of TiO$_2$.
These dopants have been considered before within DFT, often with
the generalized gradient approximation (GGA) for the
exchange-correlation functional
%(e.g.Perdew-Burke-Ernzerhof (PBE) functional~\cite{perdew96})
or within the DFT+U method and with a focus on neutral impurities
and their electronic properties.\cite{scanlon2009,
osorio-guillen08, elsasser, deak11, huy2012, robertson, celik2012,
long11cpl, long11prb, chendm10, kamisaka09, Pan2015}
%where an on-site atomic potential $U$ is introduced to account for the
%strong correlation of the Ti d-electrons.~\cite{Morgan, Watson and osorio-guillen08}
%However both PBE and DFT+U severely
%underestimate the band gap and might therefore affect the results
%obtained for the impurities. On the contrary, the HSE functional
%is known to yield reliable results for impurity doping in
%semiconductors. Yamamoto et al. ~\cite{Yamamoto} found Nb- and
%Ta-doped TiO$_2$ to be metallic , whereas Deak et al. identified
%Nb as a shallow donor.~\cite{Deak}. A similar result was obtained
%by Lee and Robertson based on the screened exchange hybrid
%functional.~\cite{Robertson} The optical absorption spectrum of
%Nb, Ta, Mo, and W was investigated with the HSE functional by Long
%and English. ~\cite{Long}.
%Although the single-particle band
%structure provides useful insight into the dopants electronic
%properties, a more complete picture emerges from a general
However, whether an impurity becomes electrically active or rather
remains inactive or even behaves amphoteric often depends on the
growth conditions (e.g. the presence of native defects) and cannot
directly be inferred from a DFT calculation on a neutral defect
alone. A more complete picture emerges from a general
thermodynamic analysis based on the defect formation
energy.\cite{freysoldt14, vdw} Such an analysis was applied by
Osorio-Guill\'en et al. on the V-b TM impurities in TiO$_2$ based
on a GGA functional together with a correction scheme that
accounts for the finite size of the supercell and the
underestimation of the GGA band gap\cite{osorio-guillen08} and
for Nb-doped TiO$_2$ a similar approach was followed in
refs.~\citenum{deak11,elsasser}. In the present work we
consider both V-b and VI-b impurities and avoid an \emph{a
posteriori} correction of the band gap by using the
Heyd-Scuseria-Ernzerhof (HSE) hybrid functional.\cite{heyd03,
heyd06}

Our work provides insight into several experimental observations.
It shows that Nb, Ta and W are electron donors leading to
transparent conducting TiO$_2$ but with different charge carrier
concentrations due to the difference in defect formation enthalpy.
In contrast, V will not easily donate electrons since the
ionization energy is too large. Moreover we demonstrate that,
depending on the actual experimental conditions, Cr acts as an
amphoteric impurity and that Mo can exist as a single or double
electron donor.

%\emph{ab initio}
% using a hybrid density functional to obtain
%accurate band gap values. These accurate band gap values are
%necessary for the calculation of accurate defect formation
%energies and thermodynamic transition levels of (charged)
%impurities in semiconductors ~\cite{vdw,Hemant}. Our study
%is an extension of previous \emph{ab initio} studies of TM-doped
%TiO$_2$.~\cite{osorio-guillen08}

The paper is organized as follows. In section \ref{sec:methods} we
present the computational details for the DFT calculations
(subsection \ref{ss:comp}) and the methodology used to calculate
the defect formation energies (subsection \ref{ss:Ef}). In section
\ref{ss:enthalpy} we present the results on the formation energy
of the different TM oxides considered in this work and in section
\ref{ss:EfResult} we discuss the defect formation energies and the
thermodynamic transition levels. Our conclusions are given in
section \ref{conclusion}.

\section{Methods\label{sec:methods}}

\subsection{Computational details\label{ss:comp}}

Our DFT calculations are performed using the projector augmented
wave method as implemented in the VASP code\cite{kresse96,
kresse99} and are spin polarized for all TM-doped systems.
We treat $3p$, $3d$ and $4s$ states as valence states for Ti, V and Cr.
Similarly, $4p$, $4d$ and $5s$ states are treated as valence states for Nb and Mo,
whereas $5p$, $5d$ and $6s$ states are treated as valence states for Ta and W.
Wave functions are expanded with a plane-wave basis set with a
cutoff energy of 400 eV.
We have used both the Perdew-Burke-Ernzerhof (PBE)
functional\cite{perdew96} within GGA and the HSE hybrid
functional\cite{heyd03, heyd06} for the exchange \& correlation
energy. More specifically we used the HSE06 functional with a
screening parameter of $\mu = 0.2$ \AA$^{-1}$ and a 25\% mixing of
Hartree-Fock exchange.

Anatase TiO$_2$ has a tetragonal structure with space group
$I4_1/amd$ (\#141). The lattice constants $a$ and $c$ and the
internal parameter $u$ of the anatase unit cell are optimized by
both PBE and HSE. A Monkhorst-Pack $k$-point mesh of $7 \times 7
\times 7$ was used for the Brillouin zone integration. The
obtained lattice constants along with the (indirect) band gap
value ($E_g$) are shown together with the experimental
values\cite{burdett87,tangh95} in Table~\ref{tab:lattice}. The
calculated lattice constants (both HSE and PBE) are in good
agreement with the experimental results, although  HSE gives a
somewhat better result. The band gap value is severely
underestimated (by $\sim$1.3 eV) in PBE, whereas  HSE
significantly improves the result and only slightly overestimates
the experimental value by $\sim$0.2 eV.
%
%\begin{table}%[H] add [H] placement to break table across pages
% \caption{Lattice constants and band gap value of anatase TiO$_2$
% obtained by PBE and HSE along with experimental values~\cite{burdett87,tangh95}.
% \label{tab:lattice}}
% \begin{ruledtabular}
% \begin{tabular}{lllll}
%     & $a$ (\AA) & $c$ (\AA) & $u$ & $E_g$ (eV) \\ \hline
% PBE & 3.811 & 9.713 & 0.207 & 2.09 \\
% HSE & 3.771 & 9.604 & 0.206 & 3.62 \\
% Experiment & 3.782 & 9.502 & 0.208 & 3.4 \\
% \end{tabular}
% \end{ruledtabular}
% \end{table}
%%

\begin{table}[tb]
\centering
\small
  \caption{\ Lattice constants and band gap value of anatase TiO$_2$
 obtained by PBE and HSE along with experimental values.\cite{burdett87,tangh95}}
  \label{tab:lattice}
  \begin{tabular}{lllll}
     & $a$ (\AA) & $c$ (\AA) & $u$ & $E_g$ (eV) \\ \hline
 PBE & 3.811 & 9.713 & 0.207 & 2.09 \\
 HSE & 3.771 & 9.604 & 0.206 & 3.62 \\
 Experiment & 3.782 & 9.502 & 0.208 & 3.4 \\
 \hline
 \end{tabular}
\end{table}
Using these optimized lattice constants we construct  $3 \times 3
\times 1$ tetragonal supercells including 108 atoms (36 Ti and 72
O atoms). Replacing one of the Ti atoms by a TM atom results in a
dopant concentration of $\sim$2.8\%. For this supercell a
Monkhorst-Pack $k$-point mesh of $2 \times 2 \times 2$ was used
for the Brillouin zone integration. For TM-doped TiO$_2$ the size
of the supercell was kept fixed but the atomic positions were
relaxed until forces were smaller than $0.05$ eV/\AA.

\subsection{Formation energies and chemical potentials\label{ss:Ef}}

The key quantity to determine the concentration, the solubility
and the thermodynamic (and optical) transition levels of an
impurity X in charge state $q$ is the defect formation energy
$E^f(X^q)$. For an impurity atom X (X = V, Cr, Nb, Ta, Mo or W)
replacing a Ti atom it is defined by\cite{vdw,freysoldt14}:
\begin{eqnarray}
% E^f(D^q) &=& E_{tot}(D^q) - E_{tot}(\mathrm{TiO_2}) + \sum_{\mathrm{X}} n_{\mathrm{X}} \mu_{\mathrm{X}}
E^{f}(X^{q}) &=& E_{tot}(X^{q}) - E_{tot}(\mathrm{TiO_2}) +
\mu_{\mathrm{Ti}}- \mu_{\mathrm{X}}
          \nonumber \\
          &&+ q(E_v + E_F + \Delta V)+\Delta E(q).
\label{eq:edefect}
\end{eqnarray}
Here, $E_{tot}(X^q)$ and $E_{tot}(\mathrm{TiO_2})$ are the total
energies of the supercell containing the defect X in charge state
$q$ and of the perfect crystal, respectively. $\mu_{\mathrm{Ti}}$
and $\mu_{\mathrm{X}}$ are the chemical potential of Ti and the
impurity X respectively. The defect formation energy depends also
on the chemical potential of the electrons (the ``Fermi energy'',
$E_F$) and is conventionally given with respect to the valence
band maximum (VBM, $E_v$) of the undoped system, where $E_F$
usually ranges from the VBM to the conduction band minimum (CBM,
$E_c$). The correction $\Delta V$ aligns the reference potential
of the undoped crystal with that of the doped one and is
calculated following the procedure introduced in ref.
\citenum{saniz13}. $\Delta E(q)$ accounts for the electrostatic
correction necessary in the case of charged defects\cite{freysoldt14}. We use the Madelung potential, as derived by
Rurali and Cartoix\`a for anisotropic systems\cite{Rurali2009},
i.e., depending on the dielectric tensor, $\varepsilon$, of the
host material. Specifically, we write $\Delta
E(q)=-\frac{q}{2}V_M(q)$, with the generalized Madelung potential
given by
\begin{eqnarray}
V_M(q)&=&\sum_{{\bf R}\neq{\bf 0}}\frac{q} {\sqrt{det\varepsilon}}
\frac{{\rm erfc} (\gamma\sqrt{{\bf R}\varepsilon^{-1}{\bf R}})}
{\sqrt{{\bf R}\varepsilon^{-1}{\bf R}}} -\frac{\pi
q}{\Omega\gamma^2}
\nonumber \\
&+&\sum_{{\bf G}\neq{\bf 0}}\frac{4\pi q}{\Omega} \frac{\exp(-{\bf
G}\varepsilon{\bf G}/4\gamma^2)} {{\bf G}\varepsilon{\bf
G}}-\frac{2\gamma q} {\sqrt{\pi det\varepsilon}}.
\end{eqnarray}
The summations of {\bf R} and {\bf G} run over the direct and
reciprocal lattices, respectively. $\gamma$ is a suitably chosen
convergence parameter and $\Omega$ is the volume of the primitive
cell. The above has proven to be an excellent correction scheme
for anisotropic systems.\cite{Rurali2009,Kumagai2014,Petretto2014} We use the
experimental values for anatase TiO$_2$,
$\varepsilon^\parallel_\infty=5.41$,
$\varepsilon^\perp_\infty=5.82$, and
$\varepsilon^\parallel_0=22.7$, $\varepsilon^\perp_0=45.1$~\cite{Gonzalez1997} Note that because of the large size of the
dielectric tensor components, the electrostatic correction is
rather small in the present case.

%Note that because of the relatively
%large value of the dielectric constants of anatase
%($\sim$40)~\cite{kim-jy06}, finite-size corrections for the
%supercell are not considered.

 The chemical potential for Ti, and the impurity X will
always be given with respect to the total energy per atom of its
bulk metallic phase, e.g. $\mu_{\mathrm{X}} = E_{tot}(\mathrm{X})
+ \Delta \mu_{\mathrm{X}}$. \footnote{For metallic Ti and X we
considered the hexagonal close-packed (space group \# 194) and
body-centered cubic (space group \# 229) structure respectively,
with the exception of Cr for which we considered the CsCl
structure with an antiferromagnetic ordering.} $\mu_{\mathrm{O}}$
is referenced to the total energy of an O atom in an O$_2$
molecule, $\mu_{\mathrm{O}} = 1/2 E_{tot}(O_2) + \Delta \mu_O$.
The chemical potentials of O and Ti should satisfy the stable
growth condition for TiO$_2$,
\begin{eqnarray}\label{eq:cptio2}
 \Delta \mu_{\mathrm{Ti}} + 2 \Delta \mu_{\mathrm{O}} = \Delta H_f
 (\mathrm{TiO_2}).
\end{eqnarray}
$\Delta H_f (\mathrm{TiO_2})$ denotes the enthalpy of formation
for TiO$_2$. Under extreme O-rich conditions, $\Delta
\mu_{\mathrm{O}} = 0$. Under extreme Ti-rich conditions, the
growth of Ti$_2$O$_3$ becomes favorable, because the Ti/O ratio in
this compound is higher than that in TiO$_2$. As a result, $\Delta
\mu_{\mathrm{Ti}}$ is bounded by the formation of Ti$_2$O$_3$,
\begin{eqnarray}
 2 \Delta \mu_{\mathrm{Ti}} + 3 \Delta \mu_{\mathrm{O}} = \Delta H_f (\mathrm{Ti_2O_3}),
\label{eq:cpti2o3}
\end{eqnarray}
where $\Delta H_f (\mathrm{Ti_2O_3})$ denotes the enthalpy of
formation for Ti$_2$O$_3$. The intersection of
Eqs.~(\ref{eq:cptio2}) and~(\ref{eq:cpti2o3}) corresponds to the
upper limit of the chemical potentials for stable TiO$_2$ growth,
i.e. $\Delta\mu_{\mathrm{Ti}}=2\Delta
H_f(\mathrm{Ti_2O_3})-3\Delta H_f(\mathrm{TiO_2})$ and
$\Delta\mu_{\mathrm{O}}=-\Delta H_f(\mathrm{Ti_2O_3})+2\Delta
H_f(\mathrm{TiO_2})$. In Table ~\ref{tab:FECPTiO} we summarize the
PBE and HSE results for the formation enthalpy of TiO$_2$ and
Ti$_2$O$_3$, and the values for $\Delta \mu_{\mathrm{Ti}}$ and
$\Delta \mu_{\mathrm{O}}$ both for O-rich and Ti-rich conditions.
The enthalpies of formation calculated by HSE are in better
agreement with the experimental values\cite{chase98} than those
calculated by PBE.
%
 %\begin{table*}%[H] add [H] placement to break table across pages
% \caption{The enthalpies of formation for TiO$_2$ and Ti$_2$O$_3$ obtained by PBE and HSE
% and the experimental values\cite{chase98}. The results obtained from a HSE calculation using the experimental structure (i.e. lattice constants and atomic positions) is given in the
% third line (HSE$^*$). The corresponding values
% for $\Delta\mu_{\mathrm{Ti}}$ and $\Delta\mu_{\mathrm{O}}$ both in O-rich and Ti-rich conditions
% are also presented. Units are in eV.\label{tab:FECPTiO}}
% \begin{ruledtabular}
% \begin{tabular}{lllllll}
% & & &\multicolumn{2}{c}{O-rich}&\multicolumn{2}{c}{Ti-rich} \\
% & $\Delta H_f (\mathrm{TiO_2})$ & $\Delta H_f (\mathrm{Ti_2O_3})$ & $\Delta \mu_{\mathrm{Ti}}$ & $\Delta \mu_{\mathrm{O}}$ &
% $\Delta \mu_{\mathrm{Ti}}$ & $\Delta \mu_{\mathrm{O}}$ \\ \hline
% PBE   & $-$9.19 & $-$14.58 & $-$9.19 & 0 & $-$1.59 & $-$3.80 \\
% HSE   & $-$9.73 & $-$15.35 & $-$9.73 & 0 & $-$1.52 & $-$4.11 \\
%% HSE with PBE lattice  & $-$9.72 & $-$15.16 & $-$9.72 & 0 & $-$1.16 & $-$4.28 \\
% HSE$^*$  & $-$9.72 & $-$15.27 & $-$9.72 & 0 & $-$1.38 & $-$4.17 \\
% Experiment & $-$9.73 & $-$15.76 & & & & \\
% \end{tabular}
% \end{ruledtabular}
% \end{table*}
%
%As a result, $\Delta \mu_{Ti}$ is bounded by the formation of Ti$_2$O$_3$ (see Fig.~\ref{fig:ChemPot}).
%As a result, $\Delta \mu_O
%= -3.80$ eV and $\Delta \mu_{Ti} = -1.59$ eV are used.
\begin{table*}[tb]
\centering
\small
  \caption{\ The enthalpies of formation for TiO$_2$ and Ti$_2$O$_3$ obtained by PBE and HSE
 and the experimental values.\cite{chase98} The results obtained from a HSE calculation using the experimental structure (i.e. lattice constants and atomic positions) is given in the
 third line (HSE$^*$). The corresponding values
 for $\Delta\mu_{\mathrm{Ti}}$ and $\Delta\mu_{\mathrm{O}}$ both in O-rich and Ti-rich conditions
 are also presented. Units are in eV.}
  \label{tab:FECPTiO}
  \begin{tabular}{lllllll}
 & & &\multicolumn{2}{c}{O-rich}&\multicolumn{2}{c}{Ti-rich} \\
 & $\Delta H_f (\mathrm{TiO_2})$ & $\Delta H_f (\mathrm{Ti_2O_3})$ & $\Delta \mu_{\mathrm{Ti}}$ & $\Delta \mu_{\mathrm{O}}$ &
 $\Delta \mu_{\mathrm{Ti}}$ & $\Delta \mu_{\mathrm{O}}$ \\ \hline
 PBE   & $-$9.19 & $-$14.58 & $-$9.19 & 0 & $-$1.59 & $-$3.80 \\
 HSE   & $-$9.73 & $-$15.35 & $-$9.73 & 0 & $-$1.52 & $-$4.11 \\
% HSE with PBE lattice  & $-$9.72 & $-$15.16 & $-$9.72 & 0 & $-$1.16 & $-$4.28 \\
 HSE$^*$  & $-$9.72 & $-$15.27 & $-$9.72 & 0 & $-$1.38 & $-$4.17 \\
 Exp. & $-$9.73 & $-$15.76 & & & & \\   \hline
 \end{tabular}
\end{table*}

The solubility limit of impurity X (X = V, Cr, Nb, Ta, Mo or W) is
determined by the formation of the corresponding oxides i.e.
%with different stoichiometries i.e.
%
\begin{eqnarray}
 x \Delta \mu_{\mathrm{X}} + y \Delta \mu_{\mathrm{O}} = \Delta H_f
 (\mathrm{X}_x\mathrm{O}_y).
 \label{eq:cpxxoy}
\end{eqnarray}
%
%Therefore we need to compute the enthalpy of formation for each
%oxide material.
%
Since HSE calculations are computationally very demanding in
particular for structural relaxations, we limit ourselves to HSE
calculations using the experimental structure (i.e. lattice
constants and atomic positions) (listed as HSE$^*$ in
Table~\ref{tab:FECPTiO}) for the formation enthalpy of the metal
oxides. As an example, we compare in Table~\ref{tab:FECPTiO} a
fully relaxed HSE calculation with one based on the experimental
structure for TiO$_2$ and Ti$_2$O$_3$. For completeness we also
added the (fully relaxed) PBE results. We notice that the fully
relaxed and HSE based on the experimental structure agree within
0.5 \% and are in good agreement with experiment. The
corresponding values for $\Delta \mu_{\mathrm{Ti}}$ and $\Delta
\mu_{\mathrm{O}}$ for both O-rich and Ti-rich conditions are
listed in Table~\ref{tab:FECPTiO}.

\section{Results\label{sec:results}}

\subsection{Formation enthalpies \& chemical potentials\label{ss:enthalpy}}

The defect formation energy defined in Eq.~(\ref{eq:edefect})
depends on the chemical potential of Ti, O and the TM impurities
and Eqs.~(\ref{eq:cptio2}-\ref{eq:cpxxoy}) provide the constraints
on these chemical potentials. In this section we present the
calculated formation enthalpies for the TM oxides and the limiting
values for the chemical potentials. For each TM we considered
several oxides with different stoichiometry as listed in
Table~\ref{tab:Hf}. As explained in the previous section we use
experimental lattice constants for the total energy calculations.
In Table~\ref{tab:LC} we summarize these constants together with
their PBE optimized values for the considered TM oxides. The
calculated enthalpies of formation are listed in
Table~\ref{tab:Hf} together with the available experimental
values. We notice that PBE and HSE calculations with the latter
using the experimental structure yield similar results and the
calculated values are in good agreement with the experimental ones
(with the exception of VO).

\begin{table*}[tb]
\centering
\small
  \caption{\ Experimental lattice constants used for the HSE energy calculations and
 the PBE optimized lattice constants. The values $a$, $b$ and $c$ are in \AA\
 and the values $\alpha$, $\beta$ and $\gamma$ are in degrees.}
  \label{tab:LC}
 % \begin{tabular*}{0.48\textwidth}{@{\extracolsep{\fill}}lll}
  \begin{tabular*}{\textwidth}{@{\extracolsep{\fill}}lllll}
 & k-points & Space group (\#) & Experiment & PBE \\ \hline
 VO & $ 14 \times 14 \times 14 $  & $Fm\overline{3}m$ (225) & $a=4.062$~$^a$ & $a=4.191$ \\
 VO$_2$ & $ 6 \times 6 \times 6 $ & $P4_2/mnm$ (136) & $a=4.5546, c=2.8514$~$^b$ & $a=4.614, c=2.799$ \\
 V$_2$O$_3$ & $ 4 \times 4 \times 4 $ & $R\overline{3}c$ (167) & $a=4.952, c=14.002$~$^c$ & $a=4.809, c=14.438$ \\
 V$_2$O$_5$ & $ 4 \times 4 \times 4 $ & $Pmmn$ (59) & $a=11.512, b=3.564, c=4.368$~$^d$ & $a=11.570, b=3.577, c=4.899$ \\
 V$_6$O$_{13}$ & $ 4 \times 4 \times 4 $ & $Pc$ (7) & $a=10.061, b=3.711, c=11.963$~$^e$ & $a=10.245, b=3.668, c=11.996$ \\
           & & & $\beta=100.93$ & $\beta=101.01$ \\ \hline
 NbO & $ 12 \times 12 \times 12 $ & $Pm\overline{3}m$ (221) & $a=4.210$~$^a$ & $a=4.260$ \\
 NbO$_2$ & $ 8 \times 8 \times 8 $ & $I4_1/a$ (88) & $a=13.71, c=5.985$~$^a$ & $a=13.936, c=6.091$ \\
 Nb$_2$O$_5$ & $ 4 \times 4 \times 4 $ & $P2/m$ (10) & $a=21.153, b=3.8233, c=19.356$~$^f$ & $a=21.652, b=3.878, c=19.815$ \\
         & & & $\beta=119.80$ & $\beta=119.86$ \\ \hline
 TaO & $ 14 \times 14 \times 14 $ & $Fm\overline{3}m$ (225) & $a=4.422$~\textit{$^{a}$} & $a=4.510$ \\
 Ta$_2$O$_5$ & $ 8 \times 8 \times 8 $ & $P6/mmm$ (191) & $a=7.248, c=3.880$~$^g$ & $a=7.346, c=3.891$ \\ \hline
 CrO$_2$ & $ 8 \times 8 \times 8 $ & $P4_2/mnm$ (136) & $a=4.421, c=2.917$~$^h$ & $a=4.456, c=2.934$ \\
 Cr$_2$O$_3$ & $ 4 \times 4 \times 4 $& $R\overline{3}c$ (167) & $a=4.961, c=13.599$~$^c$ & $a=4.949, c=13.846$ \\ \hline
% CrO$_3$ & $Ama2$ (40) & $a=4.789, b=8.557, c=5.743$ & $a=5.406, b=9.716, c=5.822$ \\ \hline
 MoO$_2$ & $ 8 \times 8 \times 8 $ & $P2_1/c$ (14) & $a=5.6109, b=4.8562, c=5.6285$~$^i$ & $a=5.639, b=4.919, c=5.694$ \\
     & & & $\beta=120.95$ & $\beta=120.74$ \\
% MoO$_3$ & $Pnma$ (62) & $a=3.9615, b=13.86625, c=3.701$~\footnotemark[10] & $a=3.973, b=16.815, c=3.705$ \\ \hline
 MoO$_3$ & $ 6 \times 6 \times 6 $ & $Pnma$ (62) & $a=3.9616, b=13.856, c=3.6978$~$^j$ & $a=3.973, b=16.815, c=3.705$ \\ \hline
 WO$_2$ & $ 4 \times 4 \times 4 $ & $P2_1/c$ (14) & $a=5.563, b=4.896, c=5.663$~$^k$ & $a=5.620, b=4.952, c=5.724$ \\
    & & & $\beta=120.47$ & $\beta=120.45$ \\
 WO$_3$ & $ 4 \times 4 \times 4 $ & $P_1$ (2) & $a=7.313, b=7.525, c=7.690$~$^l$ & $a=7.508, b=7.663, c=7.822$ \\
    & & & $\alpha=88.847, \beta=90.912, \gamma=90.940$ & $\alpha=89.4450, \beta=90.2957, \gamma=90.3133$ \\  \hline
\multicolumn{5}{l}{$^a$ Ref.~\citenum{wyckoff63}, $^b$ Ref.\citenum{newnham62}, $^c$ Ref. \citenum{newnham62}, $^d$ Ref. \citenum{enjalbert86}, $^e$
Ref. \citenum{howing03}, $^f$ Ref. \citenum{kato76}, $^g$ Ref. \citenum{terao67}, $^h$ Ref. \citenum{baur71}, $^i$ Ref. \citenum{brandt67},
$^j$ Ref. \citenum{sitepu05}, $^k$ Ref. \citenum{jiang04}, $^l$ Ref. \citenum{woodward95} }
 \end{tabular*}
\end{table*}

% \begin{table}%[H] add [H] placement to break table across pages
% \caption{Enthalpy of formation for various oxides obtained by a (fully relaxed) PBE
% and by an HSE calculation, with the latter using the experimental structure. Experimental values are also shown as a reference~\cite{chase98}.
% Units are in eV.\label{tab:Hf}}
% \begin{ruledtabular}
% \begin{tabular}{llll}
% $\Delta H_f$ & PBE & HSE & Experiment \\ \hline
% VO & $-$2.91 & $-$2.95 & $-$4.48 \\
% VO$_2$ & $-$7.03 & $-$7.00 & $-$7.37 \\
% V$_2$O$_3$ & $-$11.25 & $-$11.04 & $-$12.63 \\
% V$_2$O$_5$ & $-$16.01 & $-$16.37 & $-$16.07 \\
% V$_6$O$_{13}$ & $-$44.75 & $-$45.27 & $-$46.05 \\ \hline
% NbO & $-$2.37 & $-$1.72 & $-$4.35 \\
% NbO$_2$ & $-$7.41 & $-$7.71 & $-$8.24 \\
% Nb$_2$O$_5$ & $-$18.06 & $-$18.44 & $-$19.69 \\ \hline
% TaO & $-$1.82 & $-$1.71 & $-$ \\
% Ta$_2$O$_5$ & $-$18.26 & $-$18.94 & $-$21.20 \\ \hline
% CrO$_2$ & $-$5.78 & $-$5.87 & $-$6.20 \\
% Cr$_2$O$_3$ & $-$9.97 & $-$12.09 & $-$11.76 \\ \hline
%% CrO$_3$ & $-$6.56 & $-$1.57 & $-$6.00 \\ \hline
% MoO$_2$ & $-$5.72 & $-$5.58 & $-$6.09 \\
% MoO$_3$ & $-$7.48 & $-$6.70 & $-$7.72 \\ \hline
% WO$_2$ & $-$5.79 & $-$5.71 & $-$6.11 \\
% WO$_3$ & $-$8.67 & $-$8.21 & $-$8.74 \\
% \end{tabular}
% \end{ruledtabular}
% \end{table}
\begin{table}[tb]
\centering
\small
  \caption{\ Enthalpy of formation for various oxides obtained by a (fully relaxed) PBE
 and by an HSE calculation, with the latter using the experimental structure. Experimental values are also shown as a reference.\cite{chase98}
 Units are in eV.}
  \label{tab:Hf}
 % \begin{tabular*}{0.48\textwidth}{@{\extracolsep{\fill}}lll}
  \begin{tabular}{llll}
 $\Delta H_f$ & PBE & HSE & Experiment \\ \hline
 VO & $-$2.91 & $-$2.95 & $-$4.48 \\
 VO$_2$ & $-$7.03 & $-$7.00 & $-$7.37 \\
 V$_2$O$_3$ & $-$11.25 & $-$11.04 & $-$12.63 \\
 V$_2$O$_5$ & $-$16.01 & $-$16.37 & $-$16.07 \\
 V$_6$O$_{13}$ & $-$44.75 & $-$45.27 & $-$46.05 \\ \hline
 NbO & $-$3.95 & $-$4.16 & $-$4.35 \\
 NbO$_2$ & $-$7.41 & $-$7.71 & $-$8.24 \\
 Nb$_2$O$_5$ & $-$18.06 & $-$18.44 & $-$19.69 \\ \hline
 TaO & $-$1.82 & $-$1.71 & $-$ \\
 Ta$_2$O$_5$ & $-$18.26 & $-$18.94 & $-$21.20 \\ \hline
 CrO$_2$ & $-$5.78 & $-$5.87 & $-$6.20 \\
 Cr$_2$O$_3$ & $-$9.97 & $-$12.09 & $-$11.76 \\ \hline
% CrO$_3$ & $-$6.56 & $-$1.57 & $-$6.00 \\ \hline
 MoO$_2$ & $-$5.72 & $-$5.58 & $-$6.09 \\
 MoO$_3$ & $-$7.48 & $-$6.70 & $-$7.72 \\ \hline
 WO$_2$ & $-$5.79 & $-$5.71 & $-$6.11 \\
 WO$_3$ & $-$8.67 & $-$8.21 & $-$8.74 \\ \hline
 \end{tabular}
\end{table}

\begin{figure*}[htbp]
\centering
\includegraphics[width=1.8\columnwidth]{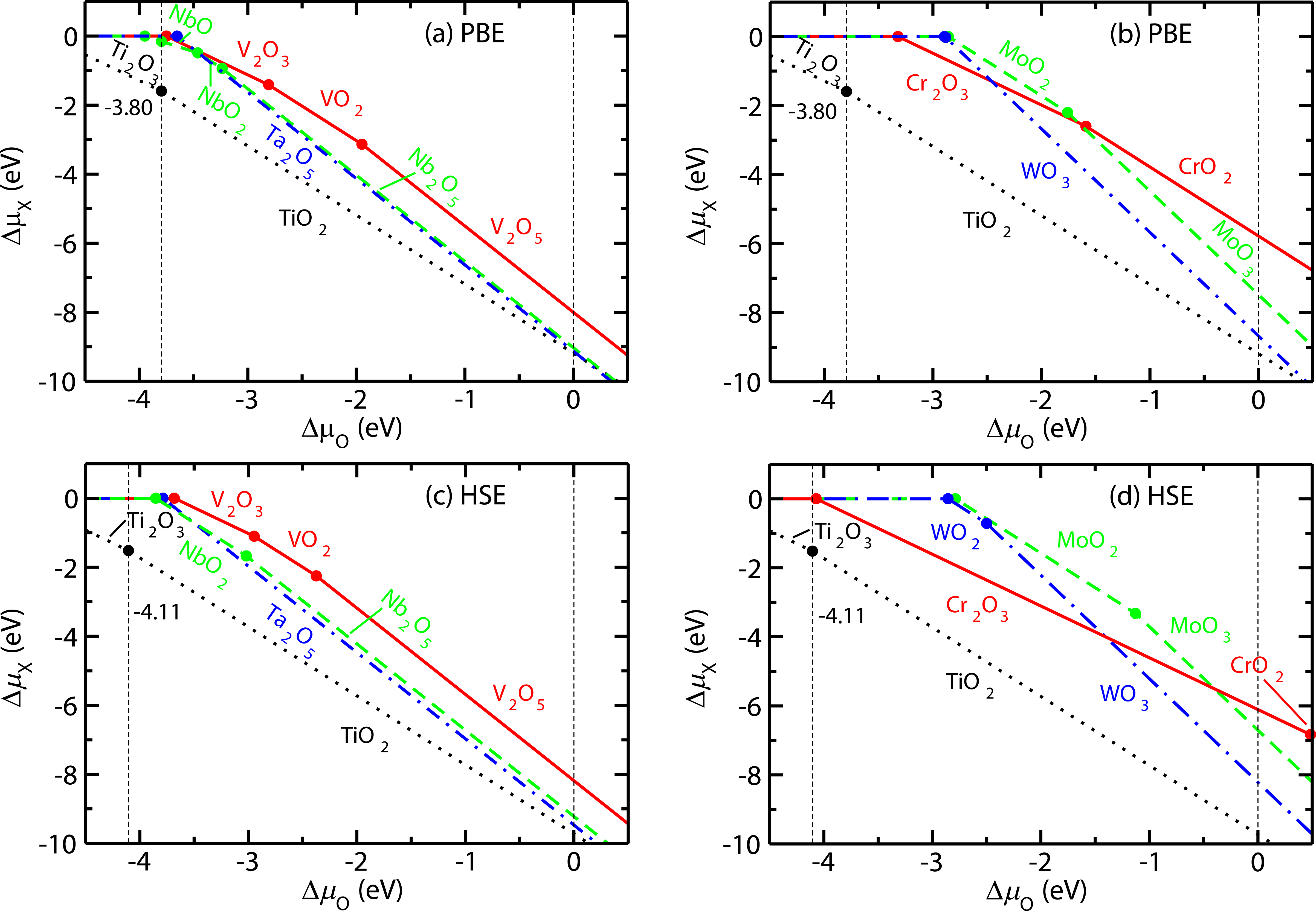}
\caption{Chemical potentials for group V-b (left side) and VI-b
(right side) elements. The upper panels are calculated within PBE,
whereas the lower panels are calculated within HSE. The vertical
dashed line at $\Delta\mu_{\mathrm{O}}=0$ corresponds to O-rich
conditions, while the other vertical dashed line at
$\Delta\mu_{\mathrm{O}}= -3.80$ eV (PBE) or $-$4.11 eV (HSE)
corresponds to Ti-rich conditions.\label{fig:cp}}
\end{figure*}

Using these enthalpies of formation for various oxide materials,
chemical potentials are computed and plotted in Fig.~\ref{fig:cp}
for (a) group V-b within PBE, (b) group VI-b within PBE, (c) group
V-b within HSE and (d) group VI-b within HSE. The chemical
potential ranges are determined by $\Delta \mu_{\mathrm{O}}=0$
(corresponding to O-rich conditions) and $\Delta
\mu_{\mathrm{O}}=-3.80$ eV (PBE) or $-4.11$ eV (HSE),
corresponding to Ti-rich conditions (i.e. O-poor conditions). In
all cases (except Nb), we find that under Ti-rich conditions
$\Delta\mu_{\mathrm{X}}=0$, i.e. the chemical potential
corresponds to that of the elemental solid phase of X (for all X).
From Fig.~\ref{fig:cp} we see that, under oxygen-rich conditions,
the chemical potentials for V, Nb and Ta are determined by
V$_2$O$_5$, Nb$_2$O$_5$ and Ta$_2$O$_5$, respectively. For group
VI-b elements, the chemical potentials for Cr, Mo and W are taken
from CrO$_2$, MoO$_3$ and WO$_3$, respectively, for PBE, whereas
they are taken from Cr$_2$O$_3$, MoO$_3$ and WO$_3$, respectively,
for HSE.

Finally, the values of $\mu_{\mathrm{X}}$ and $\mu_{\mathrm{O}}$
used for the formation energy calculation under O-rich and Ti-rich
conditions both within PBE and HSE are summarized in
Table~\ref{tab:mux}. These results will be used in the next
section where we present the formation energies of the different
TM impurity atoms.

 \begin{table*}%[H] add [H] placement to break table across pages
\centering
 \small
 \caption{\ Calculated (both in PBE and HSE) values for $\Delta\mu_{\mathrm{X}}$ and
 $\Delta\mu_{\mathrm{O}}$
 both in O-rich and Ti-rich conditions. Units are in eV.}
 \label{tab:mux}
 \begin{tabular}{lllll|llll}
 & \multicolumn{4}{c|}{PBE} & \multicolumn{4}{c}{HSE} \\
 & \multicolumn{2}{c}{O-rich} & \multicolumn{2}{c|}{Ti-rich} & \multicolumn{2}{c}{O-rich} & \multicolumn{2}{c}{Ti-rich} \\
 X & $\Delta \mu_{\mathrm{X}}$ & $\Delta \mu_{\mathrm{O}}$ & $\Delta \mu_{\mathrm{X}}$ & $\Delta \mu_{\mathrm{O}}$ & $\Delta \mu_{\mathrm{X}}$
 & $\Delta \mu_{\mathrm{O}}$ & $\Delta \mu_{\mathrm{X}}$ & $\Delta \mu_{\mathrm{O}}$ \\ \hline
 V  & $-$8.00 & 0 & 0 & $-$3.80 & $-$8.18 & 0 & 0 & $-$4.11 \\
 Nb & $-$9.03 & 0 & $-$0.15 & $-$3.80 & $-$9.22 & 0 & $-$0.05 & $-$4.11 \\
 Ta & $-$9.13 & 0 & 0 & $-$3.80 & $-$9.47 & 0 & 0 & $-$4.11 \\
 Cr & $-$5.78 & 0 & 0 & $-$3.80 & $-$5.87 & 0 & 0 & $-$4.11 \\
 Mo & $-$7.48 & 0 & 0 & $-$3.80 & $-$6.70 & 0 & 0 & $-$4.11 \\
 W  & $-$8.67 & 0 & 0 & $-$3.80 & $-$8.21 & 0 & 0 & $-$4.11 \\ \hline
 \end{tabular}
 \end{table*}
\
\subsection{Defect formation energies\label{ss:EfResult}}

\begin{figure*}
\centering
\includegraphics[width=1.8\columnwidth]{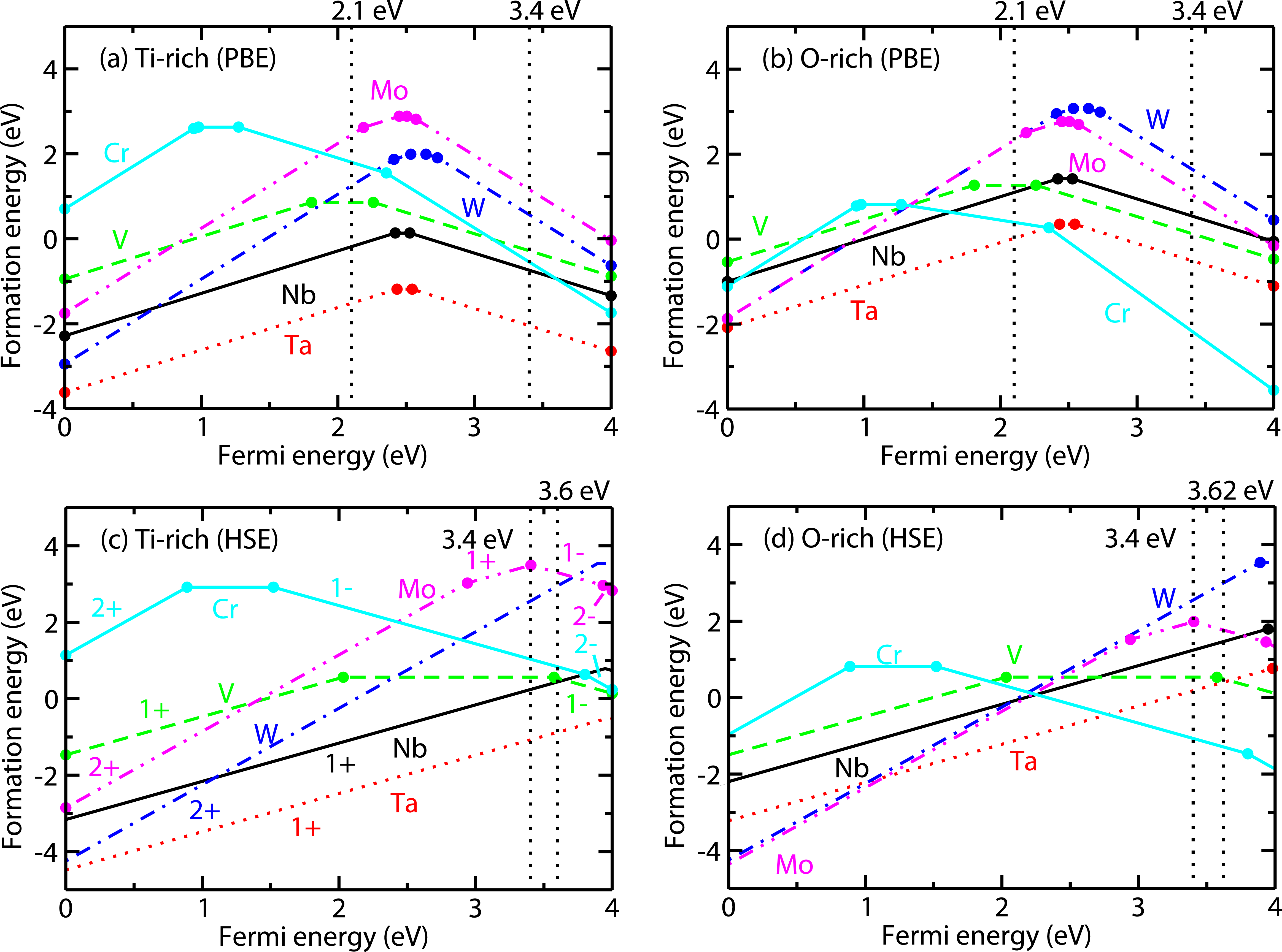}
\caption{Formation energies for TiO$_2$:X (X=V, Nb, Ta, Cr, Mo, W)
under Ti-rich (left) and O-rich (right) conditions calculated by
PBE (top) and HSE (bottom). The slope of the lines is given by the
charge state. For the HSE result in Ti-rich conditions the charge
state is explicitly indicated for convenience. The vertical dotted
bars denote the band gap positions (experimental band gap at 3.4
eV, PBE band gap at 2.1 eV and HSE band gap at 3.62 eV). Only the
lowest energy charge states are plotted and the transition levels
are marked by filled circles.\label{fig:Ef}}
%\end{figure*}
\end{figure*}

To see whether a dopant can be easily incorporated into TiO$_2$
and act as a potential donor or acceptor of electrons, we have
calculated the defect formation energies and the (thermodynamic)
transition levels for TiO$_2$:X (X=V, Nb, Ta, Cr, Mo, W) with the
HSE functional. The transition level $\epsilon(q/q')$ between two
charge states $q$ and $q'$ is defined as the Fermi energy at which
$E^f(X^q)=E^f(X^{q'})$ i.e.
\begin{eqnarray}
\epsilon(q/q')=\frac{E^{f}(X^{q};E_F=0)-E^{f}(X^{q'};E_F=0)}{q'-q}.
\label{eq:etranslevel}
\end{eqnarray}
When the Fermi level is below $\epsilon(q/q')$ charge state $q$ is
stable, otherwise charge state $q'$ is stable \footnote{a charge
state $q$ refers to a state relative to Ti$^{4+}$}. Shallow donor
(acceptor) levels correspond to transition levels that lie within
$k_BT$ from the CBM (VBM), otherwise they are labelled as deep.
The results are shown in Fig.~\ref{fig:Ef}. For a given value of
the Fermi level only the lowest energy charge state is shown for
each dopant.\footnote{For completeness we have tested the effect
of spin-orbit coupling in the case of W using the PBE functional.
As can be seen in Fig. 1, only WO$_3$ needs to be considered
(oxygen rich conditions). Spin-orbit coupling lowers $\Delta
H_f(\mathrm{W}\mathrm{O}_3)$ by $\sim$0.7 eV, so the W defect
formation energies will rise by the same amount. Note, however
that the transition levels do not depend on the chemical
potentials.}

The HSE results of Fig.~\ref{fig:Ef} show that the transition
levels $\epsilon(+/0)$ for Nb and Ta are located in the conduction
band i.e. for all Fermi level values in the band gap the preferred
charge state is $1+$, thereby demonstrating the shallow character
of the donor impurities both providing one electron to the
conduction band.\cite{vdw,freysoldt14} A similar result was
obtained for TiO$_2$:Nb by K\"orner and Els\"asser within a
LDA+SIC approach.\cite{elsasser} The result is also in line with
the metallic conductivity observed in
TiO$_2$:Nb\cite{furubayashi05,furubayashi06} and
TiO$_2$:Ta\cite{hitosugi05,Mazzolini2015}. Moreover a defect level
of $20.79$ meV below the CBM has been observed for
TiO$_2$:Nb\cite{bae}, which confirms the shallow behavior of the
impurity. The formation energy of a charged Ta impurity is lower
than that of a Nb one. This results in a higher concentration of
the former, thereby favoring a higher conductivity for TiO$_2$:Ta,
though the final value for the conductivity will also depend on
other factors such as the presence of compensating native defects.
A similar result was suggested on the basis of formation energy
calculations of neutral impurities.\cite{huy2012} In contrast to
Nb and Ta, substitutional V acts as a deep donor with the
$\epsilon(+/0)$ transition level at $1.59$ eV below the HSE CBM
(or $1.37$ eV below the experimental CBM). This result is in line
with x-ray absorption spectroscopy results\cite{Rossi2016} and EPR
measurements\cite{Tian2009} which show that vanadium is in the
charge state "V$^{4+}$" (i.e. $q=0$ in our notation), which
confirms its deep character as an electron donor. We also notice
that resonant photoemission experiments on rutile TiO$_2$:V locate
an occupied donor level $0.8$ eV below the CBM (where the band gap
value was taken as 3.0 eV).\cite{egdell-01} Our calculations
suggest that V demonstrates amphoteric behavior i.e. at Fermi
level values close to the HSE CBM it turns into an acceptor with
charge state $1-$. However, this is more likely to be an artefact
resulting from the slight overestimation of the band gap by HSE.
The results for Nb, Ta, and V are in line with those of
ref.~\citenum{osorio-guillen08}: our donor defect level for
vanadium is somewhat larger ($1.59$ versus $1.36$ eV) and our
formation energies are in general somewhat smaller.

The values of the chemical potentials appearing in
Eq.~\ref{eq:edefect} reflect the influence of the experimental
growth conditions. Using the values of Table~\ref{tab:mux} we
consider two limiting cases in Fig.~\ref{fig:Ef}, namely, O rich
and Ti rich (i.e. O poor). Comparing these conditions shows that
Nb- and Ta-doping will be more effective under O-poor conditions
i.e. under a reducing deposition
atmosphere\cite{Mazzolini2015,robertson}, whereas V doping is
hardly affected by the experimental growth conditions. For
completeness the PBE results are also given in Fig.~\ref{fig:Ef}.
Although they give qualitatively the same result for the V-b TM
atoms (using the PBE band gap), the V dopant turns out to be
considerably more shallow than in the HSE case. One also observes
that a straightforward use of the PBE transition energies with the
experimental band gap does not lead to correct results, as
expected, and a correction scheme should be invoked as in
ref.~\citenum{osorio-guillen08}.

In the case of W doping, the $2+$ charge state has the lowest
formation energy for Fermi energies between the VBM and the CBM,
thereby demonstrating its shallow donor character, which is fully
in agreement with the experimental observation that W doping
drastically decreases the resistivity.\cite{chendm10, takeuchi10}
X-ray photoelectron spectroscopy (XPS) experiments indicate that
$88$\% of the available W impurities donate two electrons to the
TiO$_2$ host system, whereas for the other $12$\% the electron
remains with the dopant atom.\cite{chendm10} However, despite the
fact that W provides two rather than one electron, the observed
conductivity of TiO$_2$:W is lower than that of TiO$_2$:Nb or
TiO$_2$:Ta.\cite{takeuchi10} The considerable higher formation
energy of the W $2+$ state compared to the $1+$ state of Nb or Ta
at Fermi energies close to the CBM (a situation that corresponds
to that of an n-doped system) is consistent with these
experimental observations, though other factors such as the
presence of electron trapping centers due to native defects (e.g.
vacancies) can play an important role too.\cite{takeuchi10}

For Mo doping the situation is different from that of W. The $2+$
state is stable for Fermi energies up to an energy of $0.68$ eV
below the HSE CBM ($0.46$ eV below the experimental CBM) from
where the $1+$ becomes stable. The HSE calculations subsequently
predict a transition to the $1-$ charge state at $0.22$ eV below
the CBM. This shows that Mo is a so-called "negative U" impurity,
where the neutral charge state is never stable and a donor turns
into an acceptor depending on the Fermi
energy.\cite{vdw,freysoldt14} However, due to the overestimation
of the band gap by the HSE functional, the negative U behavior is
likely to be an artefact. Therefore we consider Mo to be an
electron donor, donating one or two electrons to TiO$_2$ depending
on the actual value of the Fermi level, which is determined by the
overall charge neutrality of the system (i.e. including charged
native defects).
%but since the donor transition level $\epsilon(2+/+)$ is
%$0.46$ eV below the experimental CBM (and $0.68$ eV below the HSE
%CBM) it is unlikely that each impurity will donate two electrons
%to the conduction band just by thermal activation.
This result is reminiscent of the experimental situation. An
increased conductivity is observed in Mo-doped TiO$_2$ with the Mo
impurity occuring in two charge states : ~ 70\% of the dopants has
a charge state "Mo$^{6+}$" thereby donating two electrons and
~30\% has a charge state "Mo$^{5+}$" thereby donating one
electron.\cite{boen, wang, stengl} However, in
ref.~\citenum{mardare} only Mo$^{5+}$ was observed. The existence
of Mo$^{5+}$ is often attributed to the presence of O vacancies,
which indeed are known to act as shallow donors thereby moving the
Fermi level towards the CBM\cite{janotti}, where Mo will act as a
single electron donor. The VI-b impurity Cr is clearly amphoteric
with a stable $2+$ state for Fermi energies between the VBM and
$E_F=E_v+0.89$ eV and a stable $1-$ state from $E_F=E_v+1.52$ eV
to $E_c$, which makes it both a deep donor and a deep acceptor.
Since as-grown TiO$_2$ behaves as an n-type semiconductor due to
intrinsic defects acting as donor dopants (e.g. oxygen
vacancies)\cite{diebold, janotti}, the actual Fermi level is lying
close to the CBM. Our calculations show that for those Fermi
energies Cr dopants act as acceptors. This can explain the
observed reduction of the conductivity in Cr-doped
TiO$_2$.\cite{herrmann, nowotny} In ref.~\citenum{nowotny}
amphoteric behavior is observed in rutile TiO$_2$:Cr with a change
in the oxidation state of Cr from Cr$^{3+}$ to Cr$^{6+}$ in going
from a strongly reducing regime towards a more oxidizing regime.
In the former case Cr acts as an acceptor accepting one electron
(Cr$^{4+}$ $\rightarrow$ Cr$^{3+}$) whereas in the latter case it
donates two electrons (Cr$^{4+}$ $\rightarrow$ Cr$^{6+}$).
Although our calculations focus on anatase TiO$_2$ (not on rutile)
and do not consider any concomitant effects from native defects,
it is remarkable that our calculations yield the same conclusion.
According to Mizushima et al.\cite{mizushima} the Cr$^{3+}$, and
thus the acceptor level, is $2.7$ eV below the (rutile) conduction
band, which confirms the deep character of the impurity level in
TiO$_2$. From Fig.~\ref{fig:Ef} it follows that growth conditions
hardly affect the formation energy of W, but O-rich conditions are
preferred for both Mo and Cr dopants. For Cr the formation energy
drastically decreases under O-rich conditions. A similar effect is
observed for the formation energy of Mo, which is higher than that
of W under O-poor conditions but which becomes comparable to or is
even lower than that of W under O-rich conditions. As a reference
we also included the bare PBE results (i.e. without additional
band gap corrections) in Fig.~\ref{fig:Ef}. For W and Cr PBE
yields qualitatively similar results to HSE, but Mo would emerge
as a shallow donor providing two electrons per atom. The
transition level positions that appear within the (calculated)
band gap in for V, Cr and Mo are summarized in Table~\ref{tab:eps}
for both HSE and PBE.

%\begin{figure*}
 \begin{table}%[H] add [H] placement to break table across pages
\centering
 \small
 \caption{\ Calculated (in both PBE and HSE) transition level positions for dopant X with respect to the VBM. Units are in eV.}
 \label{tab:eps}
 \begin{tabular}{llll}
 X & $q/q^{\prime}$ & $\epsilon(q/q^{\prime})$ (PBE) & $\epsilon(q/q^{\prime})$ (HSE) \\ \hline
 V  & $(+/0)$ & 1.81 & 2.03 \\
    & $(0/-)$ &      & 3.57 \\
 Cr & $(2+/0)$ &     & 0.89 \\
    & $(2+/+)$ & 0.94 &  \\
    & $(+/0)$ & 0.98 &  \\
    & $(0/-)$ & 1.27 & 1.52 \\
 Mo & $(2+/+)$ & & 2.94 \\
    & $(+/-)$ & & 3.40 \\ \hline
 \end{tabular}
 \end{table}

%\begin{table}%[H] add [H] placement to break table across pages
%\caption{Calculated (in both PBE and HSE) formation energies at $E_F$ = 0.
%The values under both the O-rich and Ti-rich conditions are shown.\label{tab:EfatEF0}}
%\begin{ruledtabular}
%\begin{tabular}{llllll}
% & \multicolumn{4}{c}{PBE} & \multicolumn{4}{c}{HSE} \\
% & & \multicolumn{2}{c}{O-rich} & \multicolumn{2}{c}{Ti-rich} \\
% & q & $E^f$ (PBE) & $E^f$ (HSE) & $E^f$ (PBE) & $E^f$ (HSE) \\ \hline
% V  & $+$  & $-$0.58 & $-$1.54 & $-$0.99 & $-$1.51 \\
%    & $0$  &    1.27 &    0.54 &    0.86 &    0.56 \\
%    & $-$  & $-$     &    4.07 & $-$     &    4.09 \\
% Nb & $-$  & $-$1.05 & $-$2.20 & $-$2.48 & $-$3.21 \\
% Ta & $-$  & $-$2.13 & $-$3.26 & $-$3.66 & $-$4.52 \\
% Cr & $2+$ & $-$1.29 & $-$1.39 &    0.52 &    0.96 \\
%    & $0$  &    0.81 &    0.57 &    2.63 &    2.92 \\
%    & $-$  &    4.26 &    2.05 &    3.86 &    4.39 \\
% Mo & $2+$ & $-$2.05 & $-$4.54 & $-$1.94 & $-$3.03 \\
%    & $+$  & $-$     & $-$1.46 & $-$     &    0.04 \\
%    & $-$  & $-$     &    5.35 & $-$     &    6.85 \\
% W  & $2+$ & $-$2.05 & $-$4.43 & $-$3.13 & $-$4.43 \\
% \end{tabular}
% \end{ruledtabular}
% \end{table}

\section{Conclusions\label{conclusion}}

In this work we have identified the electronic nature of group V-b
(V, Nb, Ta) and group VI-b (Cr, Mo, W) impurities in TiO$_2$
combining thermodynamics and first-principles calculations.  The
defect formation energy for different charge states of the
impurity was calculated taking into account the impurity
solubility limit and the stability of TiO$_2$. Since reliable
defect formation energies depend on accurate band gap values we
used the HSE06 hybrid functional, which has proven to provide
realistic defect formation energies for charged impurities. All of
the obtained results are in agreement with the experimental
observations. Nb, Ta, W and Mo are identified as shallow donors
with W providing two electrons but with a considerable higher
formation energy than Nb and Ta that each donate one electron. The
lowest formation energy of the studied impurities is obtained for
Ta. Mo is a shallow donor donating one electron to TiO$_2$ but a
slight shift of the Fermi energy towards lower values (e.g.
through the presence of compensating acceptors) would turn Mo into
a double donor. V acts as a deep donor and Cr is amphoteric acting
as a donor or an acceptor depending on the actual value of the
Fermi energy and thus will act as an electron trapping center in
n-type as-grown TiO$_2$.

%%%END OF MAIN TEXT%%%

%The \balance command can be used to balance the columns on the final page if desired. It should be placed anywhere within the first column of the last page.

\balance

%If notes are included in your references you can change the title from 'References' to 'Notes and references' using the following command:
%\renewcommand\refname{Notes and references}
\section{Acknowledgments}
We gratefully acknowledge financial support from the
IWT-Vlaanderen through the ISIMADE project, the FWO-Vlaanderen
through projects G.0191.08 and G.0150.13, and the BOF-NOI of the
University of Antwerp. This work was carried out using the HPC
infrastructure of the University of Antwerp (CalcUA) a division of
the Flemish Supercomputer Center VSC, which is funded by the
Hercules foundation. M. M. acknowledges financial support from the
GOA project ''XANES meets ELNES'' of the University of Antwerp.

%%%REFERENCES%%%
\bibliography{tio2b} %You need to replace "rsc" on this line with the name of your .bib file

\providecommand*{\mcitethebibliography}{\thebibliography}
\csname @ifundefined\endcsname{endmcitethebibliography}
{\let\endmcitethebibliography\endthebibliography}{}
\begin{mcitethebibliography}{64}
\providecommand*{\natexlab}[1]{#1}
\providecommand*{\mciteSetBstSublistMode}[1]{}
\providecommand*{\mciteSetBstMaxWidthForm}[2]{}
\providecommand*{\mciteBstWouldAddEndPuncttrue}
  {\def\EndOfBibitem{\unskip.}}
\providecommand*{\mciteBstWouldAddEndPunctfalse}
  {\let\EndOfBibitem\relax}
\providecommand*{\mciteSetBstMidEndSepPunct}[3]{}
\providecommand*{\mciteSetBstSublistLabelBeginEnd}[3]{}
\providecommand*{\EndOfBibitem}{}
\mciteSetBstSublistMode{f}
\mciteSetBstMaxWidthForm{subitem}
{(\emph{\alph{mcitesubitemcount}})}
\mciteSetBstSublistLabelBeginEnd{\mcitemaxwidthsubitemform\space}
{\relax}{\relax}

\bibitem[Hashimoto \emph{et~al.}(2005)Hashimoto, Irie, and
  Fujishima]{hashimoto05}
K.~Hashimoto, H.~Irie and A.~Fujishima, \emph{Jpn. J. Appl. Phys.}, 2005,
  \textbf{44}, 8269\relax
\mciteBstWouldAddEndPuncttrue
\mciteSetBstMidEndSepPunct{\mcitedefaultmidpunct}
{\mcitedefaultendpunct}{\mcitedefaultseppunct}\relax
\EndOfBibitem
\bibitem[Gracia \emph{et~al.}(2004)Gracia, Holgado, Caballero, and
  Gonzalez-Elipe]{gracia04}
F.~Gracia, J.~P. Holgado, A.~Caballero and A.~R. Gonzalez-Elipe, \emph{J. Phys.
  Chem. B}, 2004, \textbf{108}, 17466\relax
\mciteBstWouldAddEndPuncttrue
\mciteSetBstMidEndSepPunct{\mcitedefaultmidpunct}
{\mcitedefaultendpunct}{\mcitedefaultseppunct}\relax
\EndOfBibitem
\bibitem[Tang \emph{et~al.}(1995)Tang, L\'evy, Berger, and Schmid]{tangh95}
H.~Tang, F.~L\'evy, H.~Berger and P.~E. Schmid, \emph{Phys. Rev. B}, 1995,
  \textbf{52}, 7771\relax
\mciteBstWouldAddEndPuncttrue
\mciteSetBstMidEndSepPunct{\mcitedefaultmidpunct}
{\mcitedefaultendpunct}{\mcitedefaultseppunct}\relax
\EndOfBibitem
\bibitem[Furubayashi \emph{et~al.}(2005)Furubayashi, Hitosugi, Yamamoto, Inaba,
  Kinoda, Hirose, Shimada, and Hasegawa]{furubayashi05}
Y.~Furubayashi, T.~Hitosugi, Y.~Yamamoto, K.~Inaba, G.~Kinoda, Y.~Hirose,
  T.~Shimada and T.~Hasegawa, \emph{Appl. Phys. Lett.}, 2005, \textbf{86},
  252101\relax
\mciteBstWouldAddEndPuncttrue
\mciteSetBstMidEndSepPunct{\mcitedefaultmidpunct}
{\mcitedefaultendpunct}{\mcitedefaultseppunct}\relax
\EndOfBibitem
\bibitem[Furubayashi \emph{et~al.}(2006)Furubayashi, Hitosugi, Yamamoto,
  Hirose, Kinoda, Inaba, Shimada, and Hasegawa]{furubayashi06}
Y.~Furubayashi, T.~Hitosugi, Y.~Yamamoto, Y.~Hirose, G.~Kinoda, K.~Inaba,
  T.~Shimada and T.~Hasegawa, \emph{Thin Solid Films}, 2006, \textbf{496},
  157\relax
\mciteBstWouldAddEndPuncttrue
\mciteSetBstMidEndSepPunct{\mcitedefaultmidpunct}
{\mcitedefaultendpunct}{\mcitedefaultseppunct}\relax
\EndOfBibitem
\bibitem[Hitosugi \emph{et~al.}(2005)Hitosugi, Furubayashi, Ueda, Itabashi,
  Inaba, Hirose, Kinoda, Yamamoto, Shimada, and Hasegawa]{hitosugi05}
T.~Hitosugi, Y.~Furubayashi, A.~Ueda, K.~Itabashi, K.~Inaba, Y.~Hirose,
  G.~Kinoda, Y.~Yamamoto, T.~Shimada and T.~Hasegawa, \emph{Jpn. J. Appl.
  Phys.}, 2005, \textbf{44}, L1063\relax
\mciteBstWouldAddEndPuncttrue
\mciteSetBstMidEndSepPunct{\mcitedefaultmidpunct}
{\mcitedefaultendpunct}{\mcitedefaultseppunct}\relax
\EndOfBibitem
\bibitem[Sathasivam \emph{et~al.}(2015)Sathasivam, Bhachu, Lu, Chadwick,
  Althabaiti, Alyoubi, Basahel, Carmalt, and Parkin]{sathasivam15}
S.~Sathasivam, D.~S. Bhachu, Y.~Lu, N.~Chadwick, S.~A. Althabaiti, A.~O.
  Alyoubi, S.~N. Basahel, C.~J. Carmalt and I.~P. Parkin, \emph{Sci. Rep.},
  2015, \textbf{5}, 10952\relax
\mciteBstWouldAddEndPuncttrue
\mciteSetBstMidEndSepPunct{\mcitedefaultmidpunct}
{\mcitedefaultendpunct}{\mcitedefaultseppunct}\relax
\EndOfBibitem
\bibitem[Matsumoto \emph{et~al.}(2001)Matsumoto, Murakami, Shono, Hasegawa,
  Fukumura, Kawasaki, Ahmet, Chikyow, Koshihara, and Koinuma]{matsumoto01}
Y.~Matsumoto, M.~Murakami, T.~Shono, T.~Hasegawa, T.~Fukumura, M.~Kawasaki,
  P.~Ahmet, T.~Chikyow, S.-y. Koshihara and H.~Koinuma, \emph{Science}, 2001,
  \textbf{291}, 854\relax
\mciteBstWouldAddEndPuncttrue
\mciteSetBstMidEndSepPunct{\mcitedefaultmidpunct}
{\mcitedefaultendpunct}{\mcitedefaultseppunct}\relax
\EndOfBibitem
\bibitem[Janisch and Spaldin(2006)]{janisch06}
R.~Janisch and N.~A. Spaldin, \emph{Phys. Rev. B}, 2006, \textbf{73},
  035201\relax
\mciteBstWouldAddEndPuncttrue
\mciteSetBstMidEndSepPunct{\mcitedefaultmidpunct}
{\mcitedefaultendpunct}{\mcitedefaultseppunct}\relax
\EndOfBibitem
\bibitem[Hong \emph{et~al.}(2004)Hong, Sakai, and Hassini]{hongnh04}
N.~H. Hong, J.~Sakai and A.~Hassini, \emph{Appl. Phys. Lett.}, 2004,
  \textbf{84}, 2602\relax
\mciteBstWouldAddEndPuncttrue
\mciteSetBstMidEndSepPunct{\mcitedefaultmidpunct}
{\mcitedefaultendpunct}{\mcitedefaultseppunct}\relax
\EndOfBibitem
\bibitem[Droubay \emph{et~al.}(2005)Droubay, Heald, Shutthanandan, Thevuthasan,
  Chambers, and Osterwalder]{droubay05}
T.~Droubay, S.~M. Heald, V.~Shutthanandan, S.~Thevuthasan, S.~A. Chambers and
  J.~Osterwalder, \emph{J. Appl. Phys.}, 2005, \textbf{97}, 046103\relax
\mciteBstWouldAddEndPuncttrue
\mciteSetBstMidEndSepPunct{\mcitedefaultmidpunct}
{\mcitedefaultendpunct}{\mcitedefaultseppunct}\relax
\EndOfBibitem
\bibitem[Stengl and Bakardjieva(2010)]{stengl}
V.~Stengl and S.~Bakardjieva, \emph{J. Phys. Chem. C}, 2010, \textbf{114},
  19308\relax
\mciteBstWouldAddEndPuncttrue
\mciteSetBstMidEndSepPunct{\mcitedefaultmidpunct}
{\mcitedefaultendpunct}{\mcitedefaultseppunct}\relax
\EndOfBibitem
\bibitem[Khan \emph{et~al.}(2014)Khan, Xu, Cao, and Liu]{khan}
M.~Khan, J.~Xu, W.~Cao and Z.-K. Liu, \emph{J. Nanosci. Nanotechnol.}, 2014,
  \textbf{14}, 6865\relax
\mciteBstWouldAddEndPuncttrue
\mciteSetBstMidEndSepPunct{\mcitedefaultmidpunct}
{\mcitedefaultendpunct}{\mcitedefaultseppunct}\relax
\EndOfBibitem
\bibitem[Wang \emph{et~al.}(2013)Wang, Bai, Sun, Jiang, and Lian]{wang}
S.~Wang, L.~N. Bai, H.~M. Sun, Q.~Jiang and J.~S. Lian, \emph{Powder Technol.},
  2013, \textbf{244}, 9\relax
\mciteBstWouldAddEndPuncttrue
\mciteSetBstMidEndSepPunct{\mcitedefaultmidpunct}
{\mcitedefaultendpunct}{\mcitedefaultseppunct}\relax
\EndOfBibitem
\bibitem[Houng \emph{et~al.}(2013)Houng, Liu, and Hung]{boen}
B.~Houng, C.~C. Liu and M.~T. Hung, \emph{Ceram. Int.}, 2013, \textbf{39},
  3669\relax
\mciteBstWouldAddEndPuncttrue
\mciteSetBstMidEndSepPunct{\mcitedefaultmidpunct}
{\mcitedefaultendpunct}{\mcitedefaultseppunct}\relax
\EndOfBibitem
\bibitem[Morgan \emph{et~al.}(2009)Morgan, Scanlon, and Watson]{scanlon2009}
B.~J. Morgan, D.~O. Scanlon and G.~W. Watson, \emph{J. Mater. Chem.}, 2009,
  \textbf{19}, 5175\relax
\mciteBstWouldAddEndPuncttrue
\mciteSetBstMidEndSepPunct{\mcitedefaultmidpunct}
{\mcitedefaultendpunct}{\mcitedefaultseppunct}\relax
\EndOfBibitem
\bibitem[Osorio-Guill\'en \emph{et~al.}(2008)Osorio-Guill\'en, Lany, and
  Zunger]{osorio-guillen08}
J.~Osorio-Guill\'en, S.~Lany and A.~Zunger, \emph{Phys. Rev. Lett.}, 2008,
  \textbf{100}, 036601\relax
\mciteBstWouldAddEndPuncttrue
\mciteSetBstMidEndSepPunct{\mcitedefaultmidpunct}
{\mcitedefaultendpunct}{\mcitedefaultseppunct}\relax
\EndOfBibitem
\bibitem[K\"orner and Els\"asser(2011)]{elsasser}
W.~K\"orner and C.~Els\"asser, \emph{Phys. Rev. B}, 2011, \textbf{83},
  205315\relax
\mciteBstWouldAddEndPuncttrue
\mciteSetBstMidEndSepPunct{\mcitedefaultmidpunct}
{\mcitedefaultendpunct}{\mcitedefaultseppunct}\relax
\EndOfBibitem
\bibitem[De\'ak \emph{et~al.}(2011)De\'ak, Aradi, and Frauenheim]{deak11}
P.~De\'ak, B.~Aradi and T.~Frauenheim, \emph{Phys. Rev. B}, 2011, \textbf{83},
  155207\relax
\mciteBstWouldAddEndPuncttrue
\mciteSetBstMidEndSepPunct{\mcitedefaultmidpunct}
{\mcitedefaultendpunct}{\mcitedefaultseppunct}\relax
\EndOfBibitem
\bibitem[Huy \emph{et~al.}(2012)Huy, Aradi, Frauenheim, and De\'ak]{huy2012}
H.~A. Huy, B.~Aradi, T.~Frauenheim and P.~De\'ak, \emph{J. Appl. Phys.}, 2012,
  \textbf{112}, 016103\relax
\mciteBstWouldAddEndPuncttrue
\mciteSetBstMidEndSepPunct{\mcitedefaultmidpunct}
{\mcitedefaultendpunct}{\mcitedefaultseppunct}\relax
\EndOfBibitem
\bibitem[Lee and Robertson(2013)]{robertson}
H.-Y. Lee and J.~Robertson, \emph{J. Appl. Phys.}, 2013, \textbf{113},
  213706\relax
\mciteBstWouldAddEndPuncttrue
\mciteSetBstMidEndSepPunct{\mcitedefaultmidpunct}
{\mcitedefaultendpunct}{\mcitedefaultseppunct}\relax
\EndOfBibitem
\bibitem[{\c{C}}elik and Mete(2012)]{celik2012}
V.~{\c{C}}elik and E.~Mete, \emph{Phys. Rev. B}, 2012, \textbf{86},
  205112\relax
\mciteBstWouldAddEndPuncttrue
\mciteSetBstMidEndSepPunct{\mcitedefaultmidpunct}
{\mcitedefaultendpunct}{\mcitedefaultseppunct}\relax
\EndOfBibitem
\bibitem[Long and English(2011)]{long11cpl}
R.~Long and N.~J. English, \emph{Chem. Phys. Lett.}, 2011, \textbf{513},
  218\relax
\mciteBstWouldAddEndPuncttrue
\mciteSetBstMidEndSepPunct{\mcitedefaultmidpunct}
{\mcitedefaultendpunct}{\mcitedefaultseppunct}\relax
\EndOfBibitem
\bibitem[Long and English(2011)]{long11prb}
R.~Long and N.~J. English, \emph{Phys. Rev. B}, 2011, \textbf{83}, 155209\relax
\mciteBstWouldAddEndPuncttrue
\mciteSetBstMidEndSepPunct{\mcitedefaultmidpunct}
{\mcitedefaultendpunct}{\mcitedefaultseppunct}\relax
\EndOfBibitem
\bibitem[Chen \emph{et~al.}(2010)Chen, Xu, Miao, Chen, Nakao, and
  Jin]{chendm10}
D.-M. Chen, G.~Xu, L.~Miao, L.-H. Chen, S.~Nakao and P.~Jin, \emph{J. Appl.
  Phys.}, 2010, \textbf{107}, 063707\relax
\mciteBstWouldAddEndPuncttrue
\mciteSetBstMidEndSepPunct{\mcitedefaultmidpunct}
{\mcitedefaultendpunct}{\mcitedefaultseppunct}\relax
\EndOfBibitem
\bibitem[Kamisaka \emph{et~al.}(2009)Kamisaka, Hitosugi, Suenaga, Hasegawa, and
  Yamashita]{kamisaka09}
H.~Kamisaka, T.~Hitosugi, T.~Suenaga, T.~Hasegawa and K.~Yamashita, \emph{J.
  Chem. Phys.}, 2009, \textbf{131}, 034702\relax
\mciteBstWouldAddEndPuncttrue
\mciteSetBstMidEndSepPunct{\mcitedefaultmidpunct}
{\mcitedefaultendpunct}{\mcitedefaultseppunct}\relax
\EndOfBibitem
\bibitem[Pan \emph{et~al.}(2015)Pan, Li, Zhao, Liu, Gong, Niu, Liu, and
  Chi]{Pan2015}
J.~Pan, C.~Li, Y.~Zhao, R.~Liu, Y.~Gong, L.~Niu, X.~Liu and B.~Chi, \emph{Chem.
  Phys. Lett.}, 2015, \textbf{628}, 43\relax
\mciteBstWouldAddEndPuncttrue
\mciteSetBstMidEndSepPunct{\mcitedefaultmidpunct}
{\mcitedefaultendpunct}{\mcitedefaultseppunct}\relax
\EndOfBibitem
\bibitem[Freysoldt \emph{et~al.}(2014)Freysoldt, Grabowski, Hickel, Neugebauer,
  Kresse, Janotti, and Van~de Walle]{freysoldt14}
C.~Freysoldt, B.~Grabowski, T.~Hickel, J.~Neugebauer, G.~Kresse, A.~Janotti and
  C.~G. Van~de Walle, \emph{Rev. Mod. Phys.}, 2014, \textbf{86}, 253\relax
\mciteBstWouldAddEndPuncttrue
\mciteSetBstMidEndSepPunct{\mcitedefaultmidpunct}
{\mcitedefaultendpunct}{\mcitedefaultseppunct}\relax
\EndOfBibitem
\bibitem[{Van de Walle} and Neugebauer(2004)]{vdw}
C.~G. {Van de Walle} and J.~Neugebauer, \emph{J. Appl. Phys.}, 2004,
  \textbf{95}, 3851\relax
\mciteBstWouldAddEndPuncttrue
\mciteSetBstMidEndSepPunct{\mcitedefaultmidpunct}
{\mcitedefaultendpunct}{\mcitedefaultseppunct}\relax
\EndOfBibitem
\bibitem[Heyd \emph{et~al.}(2003)Heyd, Scuseria, and Ernzerhof]{heyd03}
J.~Heyd, G.~E. Scuseria and M.~Ernzerhof, \emph{J. Chem. Phys.}, 2003,
  \textbf{118}, 8207\relax
\mciteBstWouldAddEndPuncttrue
\mciteSetBstMidEndSepPunct{\mcitedefaultmidpunct}
{\mcitedefaultendpunct}{\mcitedefaultseppunct}\relax
\EndOfBibitem
\bibitem[Heyd \emph{et~al.}(2006)Heyd, Scuseria, and Ernzerhof]{heyd06}
J.~Heyd, G.~E. Scuseria and M.~Ernzerhof, \emph{J. Chem. Phys.}, 2006,
  \textbf{124}, 219906\relax
\mciteBstWouldAddEndPuncttrue
\mciteSetBstMidEndSepPunct{\mcitedefaultmidpunct}
{\mcitedefaultendpunct}{\mcitedefaultseppunct}\relax
\EndOfBibitem
\bibitem[Kresse and Furthm\"uller(1996)]{kresse96}
G.~Kresse and J.~Furthm\"uller, \emph{Phys. Rev. B}, 1996, \textbf{54},
  11169\relax
\mciteBstWouldAddEndPuncttrue
\mciteSetBstMidEndSepPunct{\mcitedefaultmidpunct}
{\mcitedefaultendpunct}{\mcitedefaultseppunct}\relax
\EndOfBibitem
\bibitem[Kresse and Joubert(1999)]{kresse99}
G.~Kresse and D.~Joubert, \emph{Phys. Rev. B}, 1999, \textbf{59}, 1758\relax
\mciteBstWouldAddEndPuncttrue
\mciteSetBstMidEndSepPunct{\mcitedefaultmidpunct}
{\mcitedefaultendpunct}{\mcitedefaultseppunct}\relax
\EndOfBibitem
\bibitem[Perdew \emph{et~al.}(1996)Perdew, Burke, and Ernzerhof]{perdew96}
J.~P. Perdew, K.~Burke and M.~Ernzerhof, \emph{Phys. Rev. Lett.}, 1996,
  \textbf{77}, 3865\relax
\mciteBstWouldAddEndPuncttrue
\mciteSetBstMidEndSepPunct{\mcitedefaultmidpunct}
{\mcitedefaultendpunct}{\mcitedefaultseppunct}\relax
\EndOfBibitem
\bibitem[Burdett \emph{et~al.}(1987)Burdett, Hughbanks, Miller, {Richardson
  Jr.}, and Smith]{burdett87}
J.~K. Burdett, T.~Hughbanks, G.~J. Miller, J.~W. {Richardson Jr.} and J.~V.
  Smith, \emph{J. Am. Chem. Soc.}, 1987, \textbf{109}, 3639\relax
\mciteBstWouldAddEndPuncttrue
\mciteSetBstMidEndSepPunct{\mcitedefaultmidpunct}
{\mcitedefaultendpunct}{\mcitedefaultseppunct}\relax
\EndOfBibitem
\bibitem[Saniz \emph{et~al.}(2013)Saniz, Xu, Matsubara, Amini, Dixit, Lamoen,
  and Partoens]{saniz13}
R.~Saniz, Y.~Xu, M.~Matsubara, M.~Amini, H.~Dixit, D.~Lamoen and B.~Partoens,
  \emph{J. Phys. Chem. Solids}, 2013, \textbf{74}, 45\relax
\mciteBstWouldAddEndPuncttrue
\mciteSetBstMidEndSepPunct{\mcitedefaultmidpunct}
{\mcitedefaultendpunct}{\mcitedefaultseppunct}\relax
\EndOfBibitem
\bibitem[Rurali and Cartoix\'a(2009)]{Rurali2009}
R.~Rurali and X.~Cartoix\'a, \emph{Nano Lett.}, 2009, \textbf{9}, 975\relax
\mciteBstWouldAddEndPuncttrue
\mciteSetBstMidEndSepPunct{\mcitedefaultmidpunct}
{\mcitedefaultendpunct}{\mcitedefaultseppunct}\relax
\EndOfBibitem
\bibitem[Kumagai and Oba(2014)]{Kumagai2014}
Y.~Kumagai and F.~Oba, \emph{Phys. Rev. B}, 2014, \textbf{89}, 195205\relax
\mciteBstWouldAddEndPuncttrue
\mciteSetBstMidEndSepPunct{\mcitedefaultmidpunct}
{\mcitedefaultendpunct}{\mcitedefaultseppunct}\relax
\EndOfBibitem
\bibitem[Petretto and Bruneval(2014)]{Petretto2014}
G.~Petretto and F.~Bruneval, \emph{Phys. Rev. Appl.}, 2014, \textbf{1},
  024005\relax
\mciteBstWouldAddEndPuncttrue
\mciteSetBstMidEndSepPunct{\mcitedefaultmidpunct}
{\mcitedefaultendpunct}{\mcitedefaultseppunct}\relax
\EndOfBibitem
\bibitem[Gonzalez \emph{et~al.}(1997)Gonzalez, Zallen, and
  Berger]{Gonzalez1997}
R.~J. Gonzalez, R.~Zallen and H.~Berger, \emph{Phys. Rev. B}, 1997,
  \textbf{55}, 7014\relax
\mciteBstWouldAddEndPuncttrue
\mciteSetBstMidEndSepPunct{\mcitedefaultmidpunct}
{\mcitedefaultendpunct}{\mcitedefaultseppunct}\relax
\EndOfBibitem
\bibitem[M.~W.~Chase(1998)]{chase98}
J.~M.~W.~Chase, \emph{J. Phys. Chem. Ref. Data Monograph No. 9 NIST. JANAF
  Thermochemical Tables (1998) 1925.}, 1998, \textbf{9}, 1\relax
\mciteBstWouldAddEndPuncttrue
\mciteSetBstMidEndSepPunct{\mcitedefaultmidpunct}
{\mcitedefaultendpunct}{\mcitedefaultseppunct}\relax
\EndOfBibitem
\bibitem[Wyckoff(1963)]{wyckoff63}
R.~Wyckoff, \emph{Crystal Structures}, Wiley, 1963\relax
\mciteBstWouldAddEndPuncttrue
\mciteSetBstMidEndSepPunct{\mcitedefaultmidpunct}
{\mcitedefaultendpunct}{\mcitedefaultseppunct}\relax
\EndOfBibitem
\bibitem[Newnham and de~Haan(1962)]{newnham62}
R.~E. Newnham and Y.~M. de~Haan, \emph{Z. Kristallogr. - Cryst. Mater.}, 1962,
  \textbf{117}, 235\relax
\mciteBstWouldAddEndPuncttrue
\mciteSetBstMidEndSepPunct{\mcitedefaultmidpunct}
{\mcitedefaultendpunct}{\mcitedefaultseppunct}\relax
\EndOfBibitem
\bibitem[Enjalbert and Galy(1986)]{enjalbert86}
R.~Enjalbert and J.~Galy, \emph{Acta Crystallogr., Sect. C}, 1986, \textbf{42},
  1467\relax
\mciteBstWouldAddEndPuncttrue
\mciteSetBstMidEndSepPunct{\mcitedefaultmidpunct}
{\mcitedefaultendpunct}{\mcitedefaultseppunct}\relax
\EndOfBibitem
\bibitem[H{\"{o}}wing \emph{et~al.}(2003)H{\"{o}}wing, Gustafsson, and
  Thomas]{howing03}
J.~H{\"{o}}wing, T.~Gustafsson and J.~O. Thomas, \emph{Acta Crystallogr., Sect.
  B}, 2003, \textbf{59}, 747\relax
\mciteBstWouldAddEndPuncttrue
\mciteSetBstMidEndSepPunct{\mcitedefaultmidpunct}
{\mcitedefaultendpunct}{\mcitedefaultseppunct}\relax
\EndOfBibitem
\bibitem[Kato(1976)]{kato76}
K.~Kato, \emph{Acta Crystallogr., Sect. B}, 1976, \textbf{32}, 764\relax
\mciteBstWouldAddEndPuncttrue
\mciteSetBstMidEndSepPunct{\mcitedefaultmidpunct}
{\mcitedefaultendpunct}{\mcitedefaultseppunct}\relax
\EndOfBibitem
\bibitem[Terao(1967)]{terao67}
N.~Terao, \emph{Jpn. J. Appl. Phys.}, 1967, \textbf{6}, 21\relax
\mciteBstWouldAddEndPuncttrue
\mciteSetBstMidEndSepPunct{\mcitedefaultmidpunct}
{\mcitedefaultendpunct}{\mcitedefaultseppunct}\relax
\EndOfBibitem
\bibitem[Baur and Khan(1971)]{baur71}
W.~H. Baur and A.~A. Khan, \emph{Acta Crystallogr., Sect. B}, 1971,
  \textbf{27}, 2133\relax
\mciteBstWouldAddEndPuncttrue
\mciteSetBstMidEndSepPunct{\mcitedefaultmidpunct}
{\mcitedefaultendpunct}{\mcitedefaultseppunct}\relax
\EndOfBibitem
\bibitem[Brandt and Skapski(1967)]{brandt67}
B.~G. Brandt and A.~C. Skapski, \emph{Acta Chem. Scand.}, 1967, \textbf{21},
  661\relax
\mciteBstWouldAddEndPuncttrue
\mciteSetBstMidEndSepPunct{\mcitedefaultmidpunct}
{\mcitedefaultendpunct}{\mcitedefaultseppunct}\relax
\EndOfBibitem
\bibitem[Sitepu \emph{et~al.}(2005)Sitepu, O'Connor, and Li]{sitepu05}
H.~Sitepu, B.~H. O'Connor and D.~Li, \emph{J. Appl. Crystallogr.}, 2005,
  \textbf{38}, 158\relax
\mciteBstWouldAddEndPuncttrue
\mciteSetBstMidEndSepPunct{\mcitedefaultmidpunct}
{\mcitedefaultendpunct}{\mcitedefaultseppunct}\relax
\EndOfBibitem
\bibitem[Jiang and Spence(2004)]{jiang04}
N.~Jiang and J.~C.~H. Spence, \emph{Phys. Rev. B}, 2004, \textbf{70},
  245117\relax
\mciteBstWouldAddEndPuncttrue
\mciteSetBstMidEndSepPunct{\mcitedefaultmidpunct}
{\mcitedefaultendpunct}{\mcitedefaultseppunct}\relax
\EndOfBibitem
\bibitem[Woodward \emph{et~al.}(1995)Woodward, Sleight, and Vogt]{woodward95}
P.~Woodward, A.~Sleight and T.~Vogt, \emph{J. Phys. Chem. Solids}, 1995,
  \textbf{56}, 1305\relax
\mciteBstWouldAddEndPuncttrue
\mciteSetBstMidEndSepPunct{\mcitedefaultmidpunct}
{\mcitedefaultendpunct}{\mcitedefaultseppunct}\relax
\EndOfBibitem
\bibitem[Mazzolini \emph{et~al.}(2015)Mazzolini, Gondoni, Russo, Chrastina,
  Casari, and Bassi]{Mazzolini2015}
P.~Mazzolini, P.~Gondoni, V.~Russo, D.~Chrastina, C.~S. Casari and A.~L. Bassi,
  \emph{J. Phys. Chem. C}, 2015, \textbf{119}, 6988\relax
\mciteBstWouldAddEndPuncttrue
\mciteSetBstMidEndSepPunct{\mcitedefaultmidpunct}
{\mcitedefaultendpunct}{\mcitedefaultseppunct}\relax
\EndOfBibitem
\bibitem[Bae \emph{et~al.}(2012)Bae, Ha, Park, Chikyow, Chang, and Oh]{bae}
H.~Bae, J.-S. Ha, S.~Park, T.~Chikyow, J.~Chang and D.~Oh, \emph{J. Vac. Sci.
  Technol., B}, 2012, \textbf{30}, 050603\relax
\mciteBstWouldAddEndPuncttrue
\mciteSetBstMidEndSepPunct{\mcitedefaultmidpunct}
{\mcitedefaultendpunct}{\mcitedefaultseppunct}\relax
\EndOfBibitem
\bibitem[Rossi \emph{et~al.}(2016)Rossi, Calizzi, Cintio, Magkos, Amidani,
  Pasquini, and Boscherini]{Rossi2016}
G.~Rossi, M.~Calizzi, V.~D. Cintio, S.~Magkos, L.~Amidani, L.~Pasquini and
  F.~Boscherini, \emph{J. Phys. Chem. C}, 2016, \textbf{120}, 7457\relax
\mciteBstWouldAddEndPuncttrue
\mciteSetBstMidEndSepPunct{\mcitedefaultmidpunct}
{\mcitedefaultendpunct}{\mcitedefaultseppunct}\relax
\EndOfBibitem
\bibitem[Tian \emph{et~al.}(2009)Tian, Li, Gu, Jiang, Hu, and Zhang]{Tian2009}
B.~Tian, C.~Li, F.~Gu, H.~Jiang, Y.~Hu and J.~Zhang, \emph{Chem. Eng. J.},
  2009, \textbf{151}, 220\relax
\mciteBstWouldAddEndPuncttrue
\mciteSetBstMidEndSepPunct{\mcitedefaultmidpunct}
{\mcitedefaultendpunct}{\mcitedefaultseppunct}\relax
\EndOfBibitem
\bibitem[Morris and Egdell(2001)]{egdell-01}
D.~Morris and R.~Egdell, \emph{J. Mater. Chem.}, 2001, \textbf{11}, 3207\relax
\mciteBstWouldAddEndPuncttrue
\mciteSetBstMidEndSepPunct{\mcitedefaultmidpunct}
{\mcitedefaultendpunct}{\mcitedefaultseppunct}\relax
\EndOfBibitem
\bibitem[Takeuchi \emph{et~al.}(2010)Takeuchi, Chikamatsu, Hitosugi,
  Kumigashira, Oshima, Hirose, Shimada, and Hasegawa]{takeuchi10}
U.~Takeuchi, A.~Chikamatsu, T.~Hitosugi, H.~Kumigashira, M.~Oshima, Y.~Hirose,
  T.~Shimada and T.~Hasegawa, \emph{J. Appl. Phys.}, 2010, \textbf{107},
  023705\relax
\mciteBstWouldAddEndPuncttrue
\mciteSetBstMidEndSepPunct{\mcitedefaultmidpunct}
{\mcitedefaultendpunct}{\mcitedefaultseppunct}\relax
\EndOfBibitem
\bibitem[Mardare \emph{et~al.}(2014)Mardare, Cornei, Luca, Dobromir, Irimiciuc,
  Punga, Pui, and Adomnitei]{mardare}
D.~Mardare, N.~Cornei, D.~Luca, M.~Dobromir, S.~A. Irimiciuc, L.~Punga, A.~Pui
  and C.~Adomnitei, \emph{J. Appl. Phys.}, 2014, \textbf{115}, 213501\relax
\mciteBstWouldAddEndPuncttrue
\mciteSetBstMidEndSepPunct{\mcitedefaultmidpunct}
{\mcitedefaultendpunct}{\mcitedefaultseppunct}\relax
\EndOfBibitem
\bibitem[Janotti \emph{et~al.}(2010)Janotti, Varley, Rinke, Umezawa, Kresse,
  and {Van de Walle}]{janotti}
A.~Janotti, J.~B. Varley, P.~Rinke, N.~Umezawa, G.~Kresse and C.~G. {Van de
  Walle}, \emph{Phys. Rev. B}, 2010, \textbf{81}, 085212\relax
\mciteBstWouldAddEndPuncttrue
\mciteSetBstMidEndSepPunct{\mcitedefaultmidpunct}
{\mcitedefaultendpunct}{\mcitedefaultseppunct}\relax
\EndOfBibitem
\bibitem[Diebold(2003)]{diebold}
U.~Diebold, \emph{Surf. Sci. Rep.}, 2003, \textbf{48}, 53\relax
\mciteBstWouldAddEndPuncttrue
\mciteSetBstMidEndSepPunct{\mcitedefaultmidpunct}
{\mcitedefaultendpunct}{\mcitedefaultseppunct}\relax
\EndOfBibitem
\bibitem[Herrmann(2012)]{herrmann}
J.-M. Herrmann, \emph{New J. Chem.}, 2012, \textbf{36}, 883\relax
\mciteBstWouldAddEndPuncttrue
\mciteSetBstMidEndSepPunct{\mcitedefaultmidpunct}
{\mcitedefaultendpunct}{\mcitedefaultseppunct}\relax
\EndOfBibitem
\bibitem[Nowotny \emph{et~al.}(2016)Nowotny, Macyk, Wachsman, and
  Rahman]{nowotny}
J.~Nowotny, W.~Macyk, E.~Wachsman and K.~A. Rahman, \emph{J. Phys. Chem. C},
  2016, \textbf{120}, 3221\relax
\mciteBstWouldAddEndPuncttrue
\mciteSetBstMidEndSepPunct{\mcitedefaultmidpunct}
{\mcitedefaultendpunct}{\mcitedefaultseppunct}\relax
\EndOfBibitem
\bibitem[Mizushima \emph{et~al.}(1979)Mizushima, Tanaka, Asai, Iida, and
  Goodenough]{mizushima}
K.~Mizushima, M.~Tanaka, A.~Asai, S.~Iida and J.~B. Goodenough, \emph{J. Phys.
  Chem. Solids}, 1979, \textbf{40}, 1129\relax
\mciteBstWouldAddEndPuncttrue
\mciteSetBstMidEndSepPunct{\mcitedefaultmidpunct}
{\mcitedefaultendpunct}{\mcitedefaultseppunct}\relax
\EndOfBibitem
\end{mcitethebibliography}
\bibliographystyle{rsc} %the RSC's .bst file

\end{document}